%% file: proceedings.tex
\documentclass[12pt]{article}
\usepackage{graphicx}
\usepackage{hyperref}

\newcommand\pubnumber{}
\newcommand\pubdate{}

\usepackage{xcolor}
\usepackage[normalem]{ulem}

\hypersetup{colorlinks=true, linkcolor=blue,  anchorcolor=blue,  
citecolor=blue, filecolor=blue, menucolor=blue, pagecolor=blue,  
urlcolor=blue}

\useunder{\uline}{\ulined}{}%
\DeclareUrlCommand{\bulurl}{}

\def\institute{Department of Physics and Astronomy \\ University of Glasgow, University Avenue, Glasgow, G12 8QQ, UK }
\def\support{\footnote{Work supported by the Science and Technologies Funding Council of the United Kingdom}}
\def\Title#1{\begin{center} {\Large #1 } \end{center}}
\def\Author#1{\begin{center}{ \sc #1} \end{center}}
\def\Address#1{\begin{center}{ \it #1} \end{center}}

\newcommand\pubblock{\rightline{\begin{tabular}{l} \pubnumber\\
         \pubdate  \end{tabular}}}
\newenvironment{Abstract}{\begin{quotation}  }{\end{quotation}}
\newenvironment{Presented}{\begin{quotation} \begin{center} 
             PRESENTED AT\end{center}\bigskip 
      \begin{center}\begin{large}}{\end{large}\end{center} \end{quotation}}

\input econfmacros.tex

\begin{document}

\begin{titlepage}
\pubblock

\vfill
\Title{Differential cross-section measurements of boosted top quarks at $\sqrt{s}=13$~TeV with the ATLAS detector}
\vfill
\Author{Michael Fenton\support\\On behalf of the ATLAS Collaboration}
\Address{\institute}
\vfill
\begin{Abstract}
Differential cross-section measurements of highly boosted top quarks are presented. The dataset used has an integrated luminosity of $3.2$~fb$^{-1}$, recorded at a centre-of-mass energy of $\sqrt{s}=13$~TeV with the ATLAS detector  at the CERN Large Hadron Collider in 2015. Events are selected in the lepton + jets channel, containing one isolated lepton and a large radius jet that is identified as originating from a top quark using substructure tagging techniques. The measured transverse momentum and absolute rapidity distributions are unfolded to remove detector effects and compared to a range of Monte Carlo simulations. The transverse momentum distribution shows that all Monte Carlo generators used predict a harder spectrum than observed in data, while the rapidity distribution agrees well between MC and data.
\end{Abstract}
\vfill
\begin{Presented}
$9^{th}$ International Workshop on Top Quark Physics\\
Olomouc, Czech Republic,  September 19--23, 2016
\end{Presented}
\vfill
\end{titlepage}
\def\thefootnote{\fnsymbol{footnote}}
\setcounter{footnote}{0}

\section{Introduction}

The top quark occupies a unique position in the Standard Model (SM). It's large mass, close to the electro-weak symmetry breaking scale, means that precision top measurements are a likely window to physics beyond the SM, where effects from new physics may modify top kinematics. The ATLAS Collaboration \cite{atlas} has previously measured top quark kinematics in the lepton+jets channel \cite{8tevatlas}, and similar measurements have also been performed by CMS \cite{8tevcms}. The largest deviation from the SM expectation in these measurements was observed in the top quark $p_T$ spectra, with most Monte-Carlo (MC) simulations predicting a harder spectra than was observed.

 To maximise reach and sensitivity to the high $p_T$ region, measurements can be performed in the ``boosted'' regime. A top quark can be considered boosted if its decay products are sufficiently collimated as to be detected within the cone of a single jet, usually with a radius parameter of around 1. Since typically the radial separation of a particle decay products is approximately $R \approx 2m / p_T$, a top quark can be considered to be boosted if it is produced with around $p_T \geq 300$~GeV.  
 
 ATLAS \cite{8tevatlasboosted} and CMS \cite{8tevcmsboosted}, both performed measurements of boosted top quark $p_T$ at $\sqrt{s}=8$~TeV. The first ATLAS measurement of this spectra at $\sqrt{s}=13$~TeV, as well as the rapidity of top quark, is reported below.



\section{Event Topology + Selection}

\begin{figure}[htb]
\centering
\includegraphics[width=0.8\linewidth]{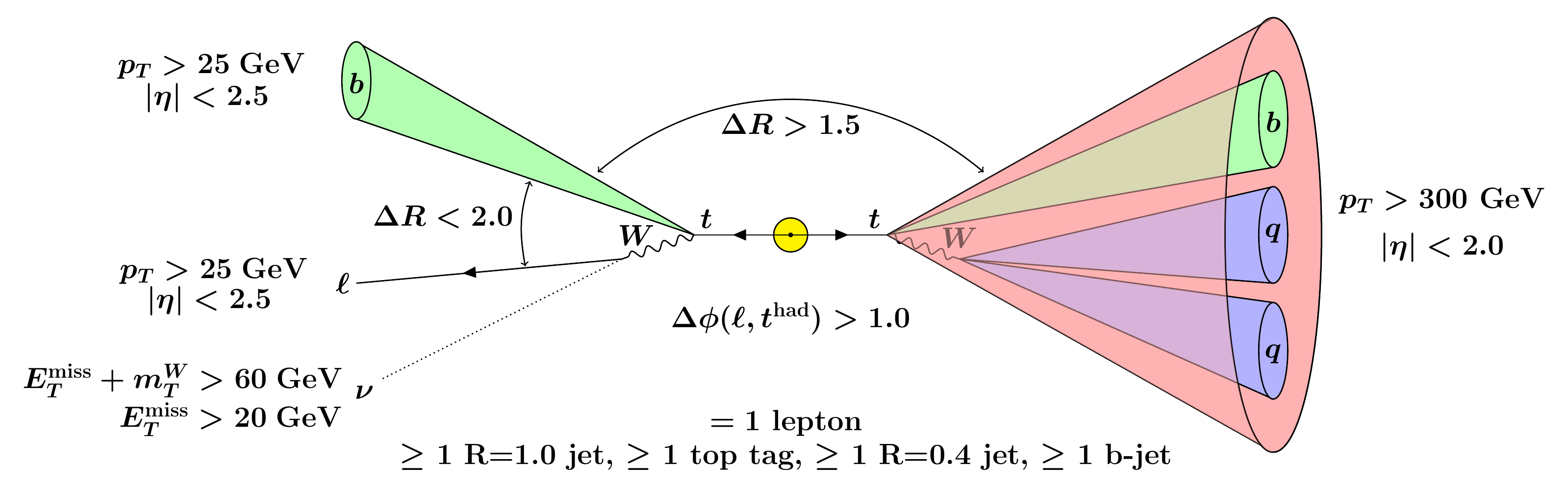}
\caption{Summary of kinematic and topological cuts used to select boosted top quark events.}
\label{fig:topology}
\end{figure}

The analysis is performed in the lepton+jets channel, whereby one top quark decays leptonically and one decays hadronically. The analysis requires exactly one lepton ($e$ or $\mu$) and at least one anti-kt R=1.0 jet. Top jets are identified using a simple tagger \cite{atlastoptagging} which uses $p_T$ dependent substructure cuts optimised on 13~TeV MC, and the leading large radius jet that passes this tagger is considered the hadronic top candidate. To further improve the signal selection, at least one b-tagged small radius jet is required in the event, either inside the leading top jet or near to the lepton. Further requirements are made on $E_T^{\rm miss}$ and $m_T^W$, as well as some angular cuts, to optimise the signal selection. The full details of these cuts are shown in Figure \ref{fig:topology}. 

The expected $t\bar{t}$ signal is predicted using a Powheg+Pythia6 MC sample. The single top background is also modelled using Powheg+Pythia6, the $t\bar{t}+V$ background is estimated with Madgraph+Pythia6, and the W+jets, Z+jets and Diboson backgrounds are modelled using Sherpa2.1. The W+jets sample has additional data-driven scale factors applied to better model the charge asymmetry and heavy flavour content of the sample, while the QCD multijet background is estimated using a fully data-driven matrix method. 

\begin{figure}[htb]
\centering
\includegraphics[width=0.4\linewidth]{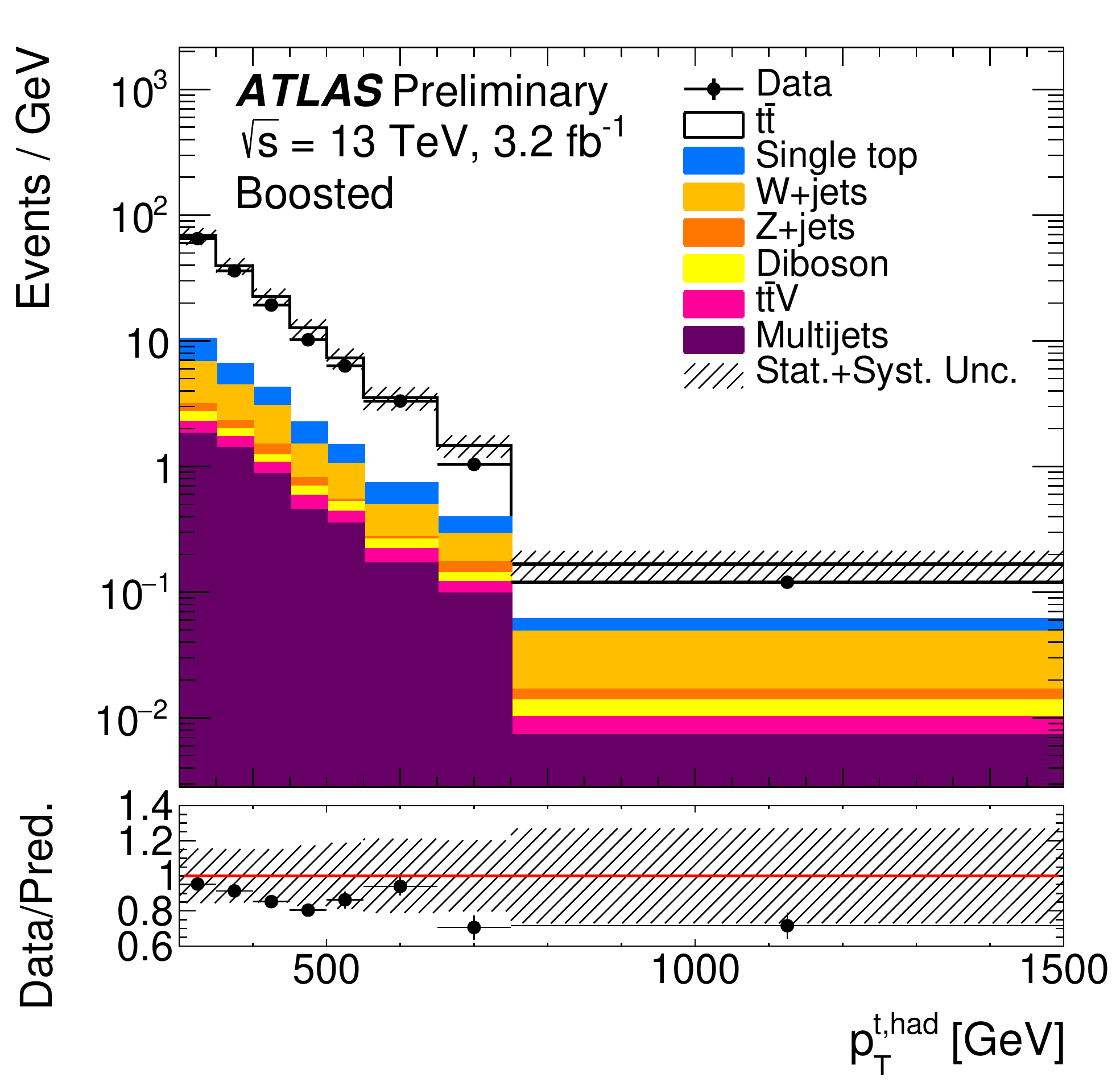}
\includegraphics[width=0.4\linewidth]{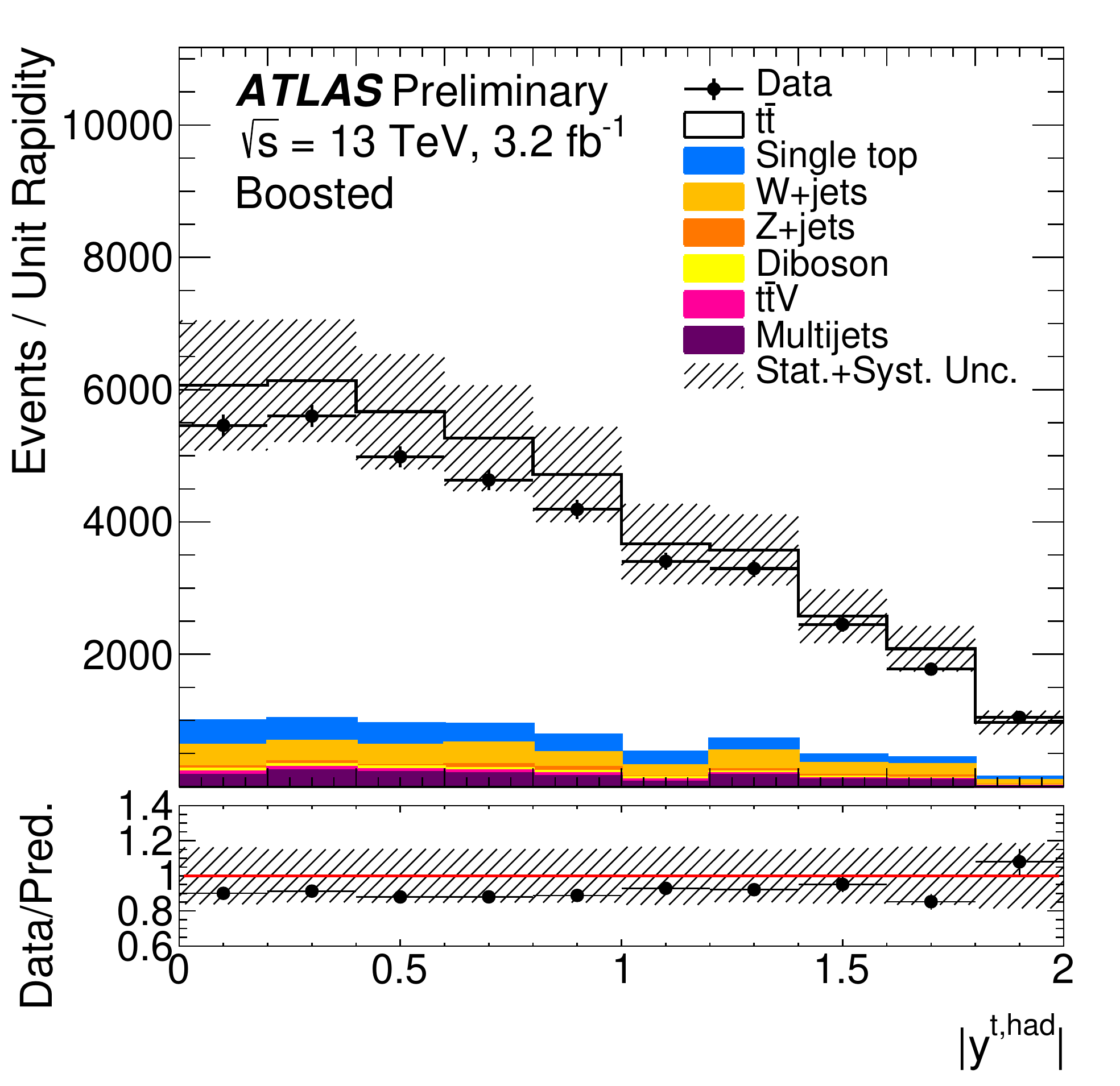}
\caption{Detector level distributions of $p_T^{t,had}$ and $|y^{t,had}|$, the two variables of interest in the analysis \cite{conf}.}
\label{fig:reco}
\end{figure}

The pre-unfolding distributions of the two variables of interest are shown in Figure \ref{fig:reco}. Good agreement is seen between prediction and data for the absolute value of the top jet rapidity, $|y^{t,had}|$, while a clear slope is present in the transverse momentum of the top jet, $p_T^{t,had}$.

\section{Unfolding Procedure}

The data is unfolded to remove detector effects to a fiducial phase space using Iterative Bayesian unfolding, with 4 iterations. This procedure involves two bin-by-bin corrections factors, $f_{\rm{eff}}$ and $f_{\rm{acc}}$, which correct for events which fail detector level and particle level selection, respectively. The expected background contribution is subtracted and a matrix $\mathcal{M}$ of bin to bin migrations is constructed. The final differential cross-section is given after normalisation by the luminosity $\mathcal{L}$ and the bin width $\Delta X$. This procedure is summarised in Equation \ref{eqn:unf}, where $\mathcal{M}_{ij}^{-1}$ represents the Bayesian unfolding procedure.

\begin{equation}
\frac{{\rm d}\sigma^{\rm fid}}{{\rm d}X^i} \equiv \frac{1}{\mathcal{L} \cdot \Delta X^i} \cdot  f_{\rm eff}^i \cdot \sum_j \mathcal{M}_{ij}^{-1} \cdot  f_{\rm acc}^j \cdot \left(N_{\rm reco}^j - N_{\rm bkg}^j\right)\hbox{}
\label{eqn:unf}
\end{equation}

The migration matrices are required to be highly diagonal in the chosen binning, such that all diagonal elements are above 50\%. 

%

\section{Uncertainties}

\begin{figure}[htb]
\centering
\includegraphics[width=0.4\linewidth]{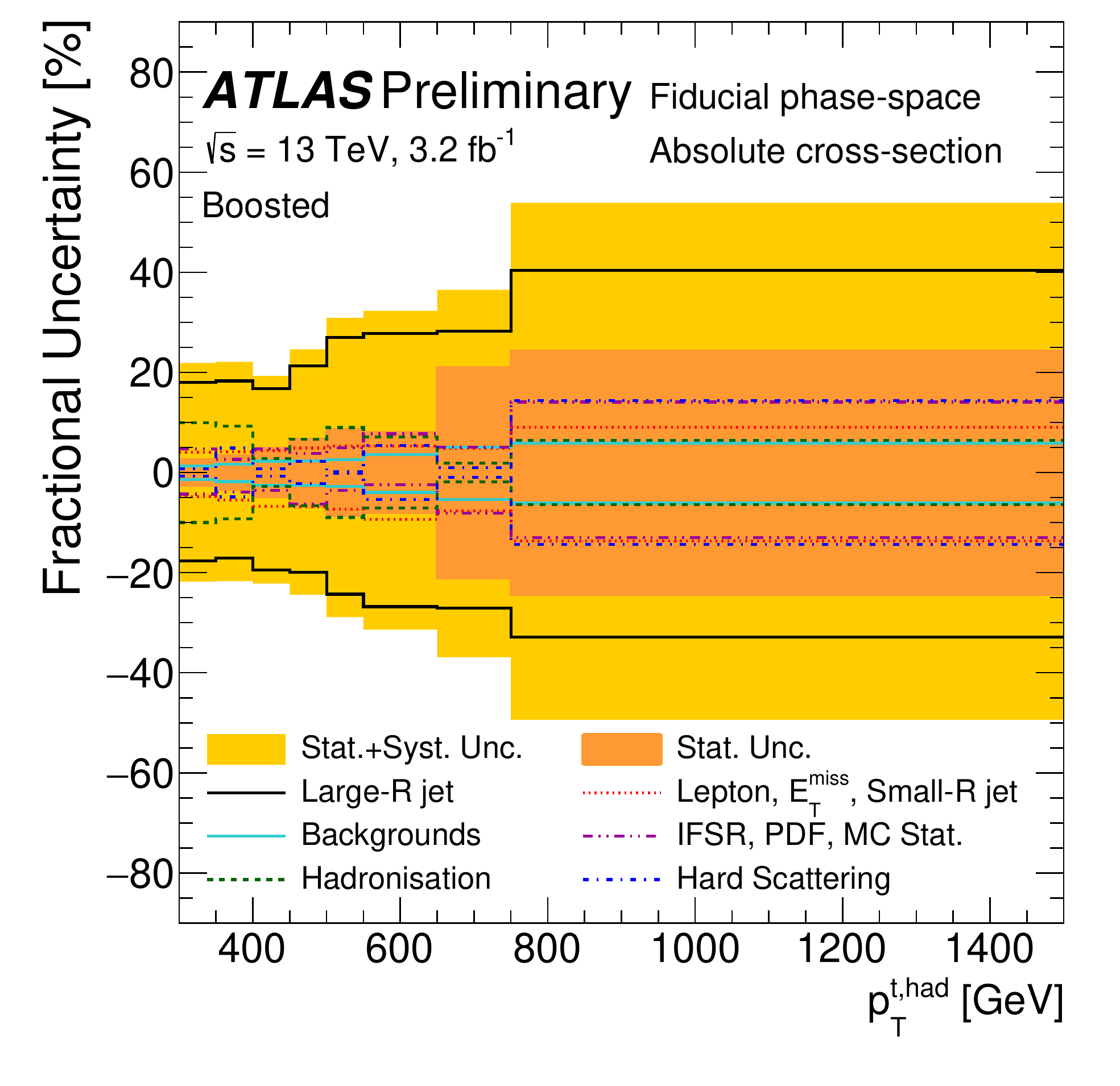}
\includegraphics[width=0.4\linewidth]{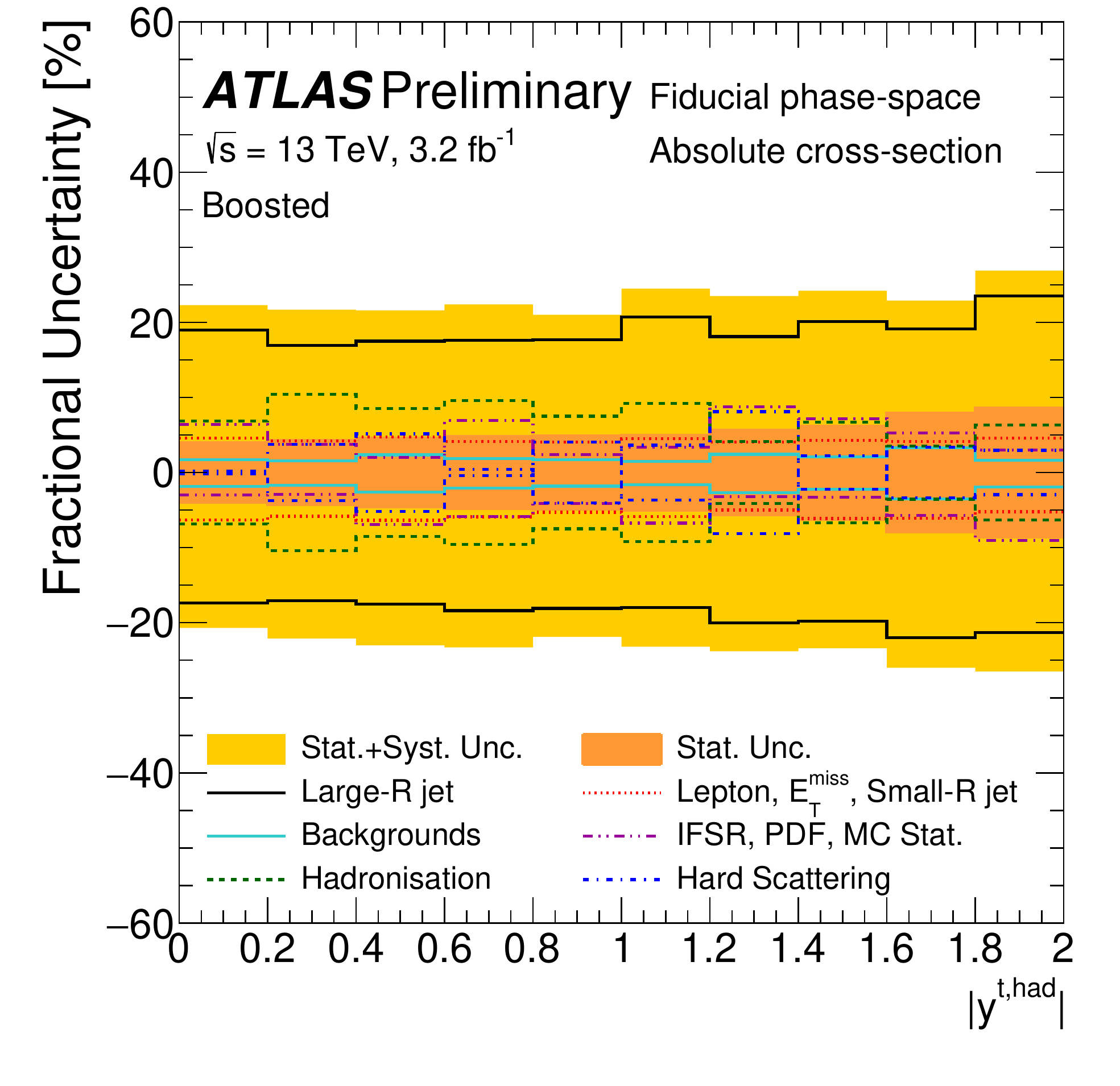}

\caption{Fractional uncertainties for transverse momentum (left) and absolute rapidity (right) \cite{conf}.}
\label{fig:unc}
\end{figure}

The measurement is limited primarily by the systematic uncertainties related to the jet energy scale, tracking, and substructure of the large-R jets, as shown for each bin of the absolute differential cross-section in Figure \ref{fig:unc}. Other significant contributions to the uncertainty come from the choice of hard scattering and hadronisation models as well as the limited statistics, particularly at high $p_T^{\rm t,had}$. Normalised differential distributions, where some systematic contributions cancel, are also included in the CONF note \cite{conf}. 

\section{Results}

\begin{figure}[htb]
\centering
\includegraphics[width=0.4\linewidth]{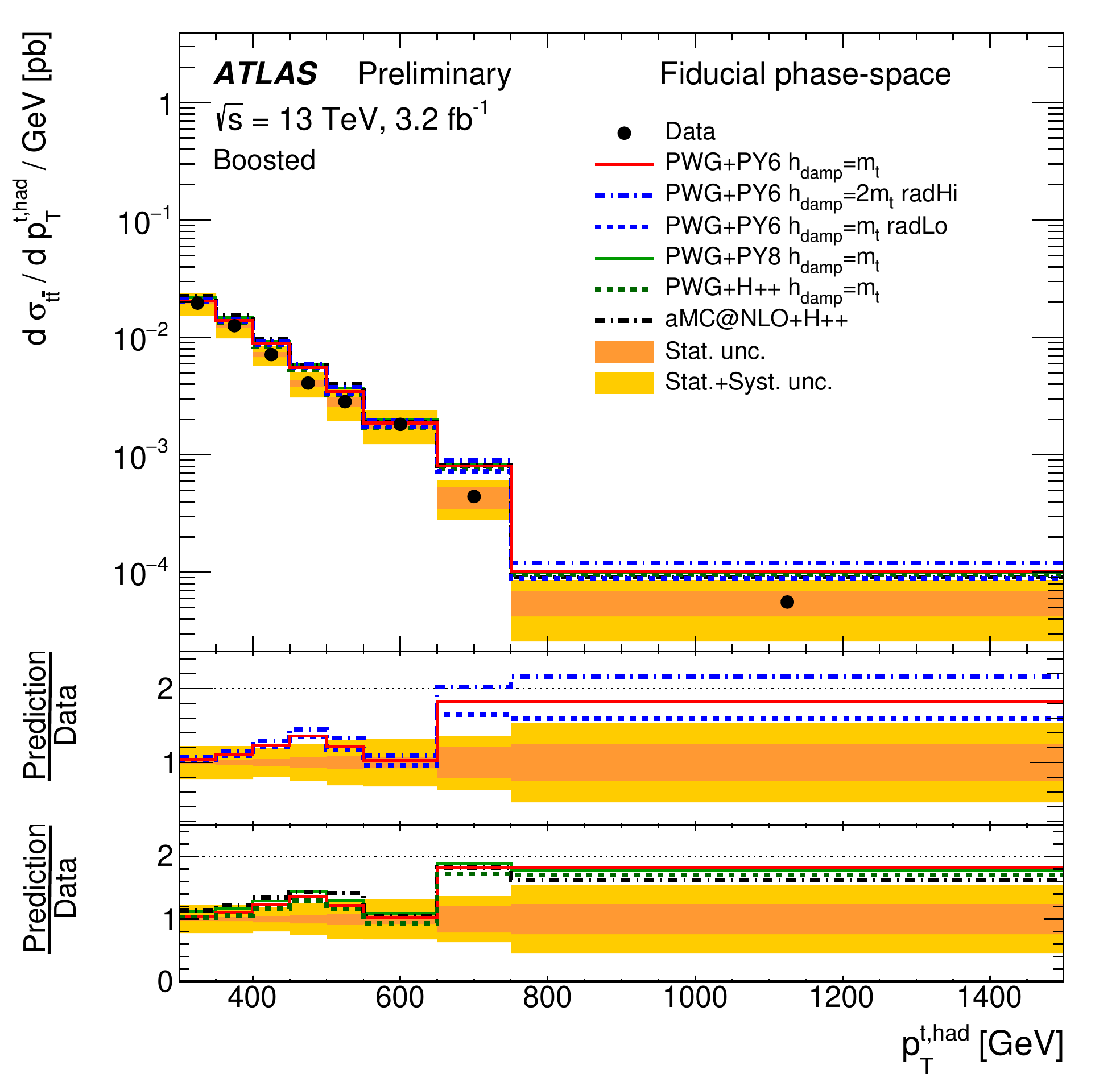}
\includegraphics[width=0.4\linewidth]{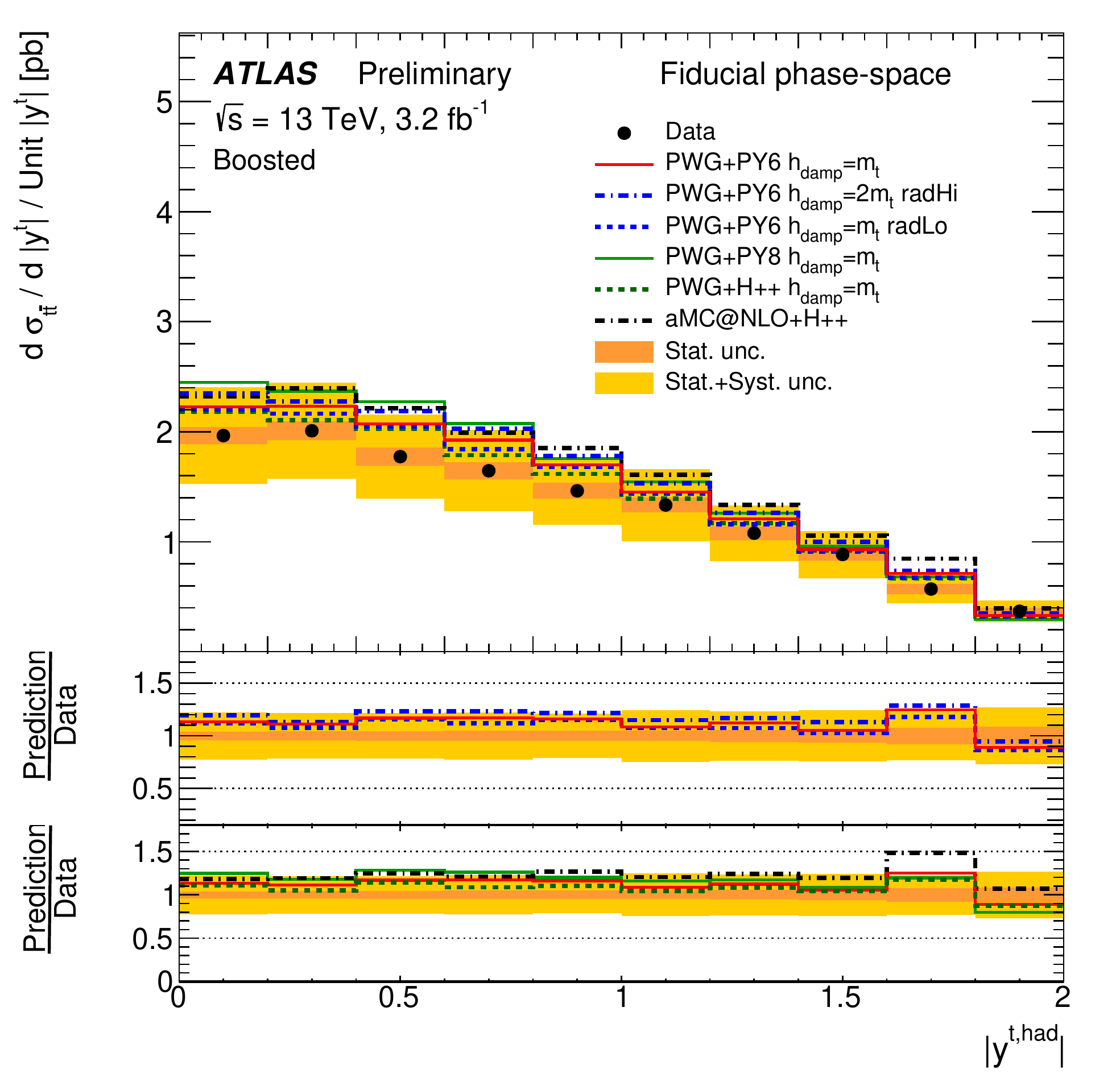}

\caption{Final results comparing unfolded data to Monte Carlo predictions transverse momentum (left) and absolute rapidity (right) \cite{conf}.}
\label{fig:results}
\end{figure}

The final unfolded data are shown and compared to multiple MC generators in Figure \ref{fig:results}. It can be seen that the $p_T^{\rm t,had}$ spectrum continues to show an increasingly large discrepancy as the $p_T$ increases, similar to that seen in all measurements in Run 1, with all predictions outwith the uncertainty bands in the final two bins. The $|y^{\rm t,had}|$ shows generally good agreement between data and MC.

\section{Conclusions}

Differential cross-sections have been measured of boosted top quarks for the first time in $\sqrt{s}=13$~TeV data with the ATLAS detector, as a function of both top quark $p_T^{\rm t,had}$ and $|y^{\rm t,had}|$. The $p_T^{\rm t,had}$ spectrum continues to show some tension with SM expectation, in line with previous measurements. The $|y^{\rm t,had}|$ distribution, previously unmeasured for boosted tops, shows broad agreement with predictions within the uncertainties. 
%
%

\end{document}

%% file: econfmacros.tex
\def\beq{\begin{equation}}
\def\eeq#1{\label{#1}\end{equation}}
\def\eeqn{\end{equation}}

\def\beqa{\begin{eqnarray}}
\def\eeqa#1{\label{#1}\end{eqnarray}}
\def\eeqan{\end{eqnarray}}

\let\bar=\overbar

\def\Dslash{\not{\hbox{\kern-4pt $D$}}}
\def\dslash{\not{\hbox{\kern-2pt $\del$}}}

\def\msb{{\bar{\ssstyle M \kern -1pt S}}}

